\documentstyle[aps,prb,epsf,twocolumn,floats]{revtex}
\begin{document}
\draft

\twocolumn[\csname @twocolumnfalse\endcsname
\widetext

\title{Quantum Critical Behavior in Kondo Systems}

\author{Qimiao Si$^{(a)}$, J. Lleweilun Smith$^{(a)}$,
and Kevin Ingersent$^{(b)}$}
\address{$^{(a)}$
Department of Physics, Rice University, Houston, TX 77251--1892 \\
$^{(b)}$Department of Physics, University of Florida, Gainesville,
FL 32611--8440}

\maketitle

\begin{abstract}
This article briefly reviews three topics related to the quantum
critical behavior of certain heavy-fermion systems. First, we summarize
an extended dynamical mean-field theory for the Kondo lattice,
which treats on an equal footing the quantum fluctuations
associated with the Kondo and RKKY couplings. The dynamical mean-field
equations describe an effective Kondo impurity model with
an additional coupling to vector bosons.
Two types of quantum phase transition appear to be possible
within this approach---the first a conventional spin-density-wave
transition, the second driven by local physics.
For the second type of transition to be realized, the effective impurity
model must have a quantum critical point exhibiting an anomalous local spin
susceptibility.
In the second part of the paper, such a critical point is shown to occur
in two variants of the Kondo impurity problem.
Finally, we propose an operational test for the existence of quantum
critical behavior driven by local physics. Neutron scattering results
suggest that CeCu$_{6-x}$Au$_x$ passes this test.

\end{abstract}

\vskip 0.2 in

]
\narrowtext

\section{Introduction}
\label{sec:intro}

\begin{figure}
\vspace{.2in}
\centering
\vbox{\epsfxsize=65mm\epsfbox{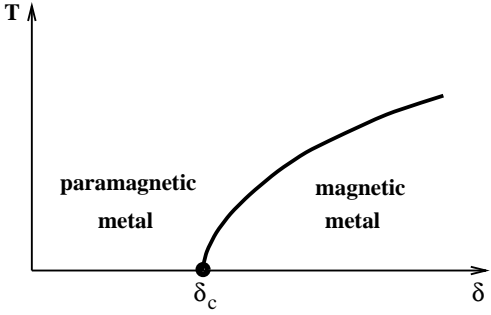}}
\vspace{1ex}
\caption{Generic phase diagram of a heavy-fermion metal
which exhibits a quantum critical point at temperature $T=0$
and at a critical value of some tuning parameter, $\delta = \delta_c$.}
\label{qpt-hf}
\end{figure}

Heavy fermions close to a quantum phase transition\cite{ITP}
represent an important class of non-Fermi liquid metals.
The transition is typically between a paramagnetic
metal and a magnetic metal.
The generic phase diagram is illustrated in Fig.~\ref{qpt-hf},
where $\delta$ represents some tuning parameter.
Non-Fermi liquid behavior occurs in the quantum critical regime
about $\delta = \delta_c$. In one class of materials, which
includes\cite{Mathur} CePd$_2$Si$_2$ and CeIn$_3$, increasing $\delta$
corresponds to decreasing external pressure.
In other cases, $\delta$ controls the stoichiometry.
For instance,\cite{Lohneysen} CeCu$_{6-x}$Au$_x$ is paramagnetic
for $x<x_c\approx 0.1$ and antiferromagnetic for $x>x_c$.

The magnetic phase transition in heavy-fermion systems is generally
thought to result from competition between Kondo and RKKY physics.
The Kondo interaction tends to quench the local moments using
the spins of the conduction electrons, whereas the RKKY interaction
promotes local-moment ordering.
The transition occurs when these two processes are of roughly equal
importance.

Historically, the interplay between Kondo and RKKY effects was first
studied in the ``Kondo necklace,'' a one-dimensional Kondo lattice
model simplified by the suppression of all charge degrees of freedom.
Using a (static) mean-field approach, Doniach\cite{Doniach}
found that increasing $I_{\text{RKKY}} / T_{\text{K}}$, the ratio of the
effective RKKY and Kondo scales, leads to a continuous transition
from a Kondo-insulator-like state to an antiferromagnetically
ordered state.
Much subsequent work has been devoted to the full Kondo lattice
model in one dimension.\cite{Ueda}
A combination of numerical and analytical methods have
yielded results very different from Doniach's:
at half-filling, a spin liquid forms for arbitrary values of
$I_{\text{RKKY}} / T_K$;
at other fillings, the system evolves from a
Luttinger liquid to a ferromagnetic metal as the
ratio $I_{\text{RKKY}} / T_K$ is {\em decreased}.
While these differences stem in part from the special properties
of interacting fermions in one dimension, they also highlight the
importance of quantum fluctuations in the Kondo lattice.

The competition between Kondo and RKKY physics has also been
studied in the two-impurity Kondo problem.\cite{Jones}
For weak antiferromagnetic RKKY interactions, single-impurity Kondo
behavior is recovered. For a very large $I_{\text{RKKY}} / T_K$, by contrast,
the two local moments lock themselves into a singlet, and no Kondo
screening takes place. Under special conditions, an unstable non-Fermi
liquid fixed point separates the two stable Fermi liquid regimes.

In order to better understand the physics of heavy-fermion metals, it is
necessary to address the dynamical competition between the Kondo and RKKY
interactions in spatial dimensionalities $D>1$. In this connection, a
recently developed extension\cite{Smith1,Smith3,Kajueter,Chitra} of the
dynamical mean-field theory (DMFT) of the large-$D$
limit\cite{Georges,Metzner} offers a promising avenue of approach.
Unlike the standard DMFT, which incorporates only purely local dynamics,
the extended DMFT takes into account nonlocal quantum fluctuations
such as might arise due to RKKY interactions.
The extended DMFT amounts\cite{Smith3} to a resummation of a
diagrammatic $1/D$ expansion.\cite{Schiller,Jarrell}

The extended DMFT represents the Kondo lattice by a self-consistently
determined impurity problem in which a single local moment couples to
vector bosons as well as to a fermionic band.
As will be explained below, the extended DMFT suggests the possibility
of two different types of quantum critical point in the Kondo lattice.
A scenario of particular interest is a transition driven
by anomalous local Kondo physics. We note that other
authors\cite{Schroder,Coleman} have also emphasized
the role played by local dynamics in heavy
fermions near quantum criticality.

The remainder of the paper is organized as follows.
Section~\ref{mean-field-theory} describes the extended DMFT
for the Kondo lattice. In Section~\ref{Kondo}, we summarize
recent work on phase transitions in related Kondo impurity
problems. Finally, Section~\ref{qpt} provides new insights
into the quantum critical behavior in heavy fermions.
In particular, we introduce a criterion which can be used to
identify quantum critical behavior driven by local physics.

\section{The extended dynamical mean-field theory}
\label{mean-field-theory}

For simplicity, we neglect valence fluctuations and take as
our starting point the Kondo lattice model:
\begin{equation}
{\cal H}
= \sum_{\langle ij\rangle,\sigma} t_{ij} c_{i\sigma}^{\dagger}c_{j\sigma}
+ \sum_i J \vec{S}_{i} \cdot \vec{s}_{c,i}
+\sum_{\langle ij\rangle} I_{ij} \vec{S}_{i} \cdot \vec{S}_{j} ,
\label{kondo-lattice}
\end{equation}
where $\vec{S}_{i}$ denotes a localized spin on site $i$,
and $\vec{s}_{c,i}$ represents the net conduction-electron spin
at site $i$.
Equation~(\ref{kondo-lattice}) includes both a local Kondo coupling $J$
and an explicit exchange coupling $I_{ij}$ between local moments.
We will consider only the nearest-neighbor terms for $I_{ij}$ and for
the electron hopping $t_{ij}$.
(Further-neighbor couplings can be easily incorporated.)

Even for $I_{ij}=0$, the Kondo coupling in Eq.~(\ref{kondo-lattice}) generates
an indirect RKKY interaction between nearest-neighbor local moments of strength
$I_{\langle ij\rangle}^{\text{ind}}\propto(J t_{\langle ij\rangle})^2$.
However, this interaction does not survive in the standard large-$D$ DMFT for
the Kondo lattice,\cite{Georges} which rescales the nearest-neighbor hopping
so that $t_{\langle ij\rangle} \rightarrow t_0/\sqrt{D}$, where $t_0$ is a
constant.
With this rescaling, $I_{\langle ij\rangle}^{\text{ind}}\propto 1/D$, which
vanishes faster than the hopping and becomes negligible for
$D\rightarrow\infty$.

The extended DMFT\cite{Smith1,Smith3,Kajueter,Chitra} ensures that the
quantum fluctuations associated with the RKKY interaction are retained at
the mean-field level by rescaling the hopping and the explicit exchange
coupling with the same power of $1/D$:
$t_{\langle ij\rangle}\rightarrow t_0/\sqrt{D}$ and
$I_{\langle ij\rangle} \rightarrow I_0/\sqrt{D}$, where $I_0$ is a constant.
This procedure is well-defined for $D=\infty$ provided
that the Hartree contribution is treated with care. In terms of diagrammatic
perturbation theory, it can be shown\cite{Smith3} that the extended DMFT is
equivalent to a conserving resummation to infinite order of a $1/D$ expansion
series.

The mean-field equations resulting from the extended DMFT can be expressed in
terms of an effective action for a single lattice site (the ``impurity'' site):
\begin{eqnarray}
S_{\text{imp}} &=& - \int_{0}^{\beta} \! d\tau \int_{0}^{\beta} \! d\tau'
\left[ \sum_{\sigma} c_{\sigma}^{\dagger}(\tau) G_0^{-1}(\tau - \tau')
c_{\sigma}(\tau')
\right. \nonumber\\ && \left.
+ \vec{S}(\tau) \cdot \chi_{0}^{-1}(\tau - \tau')
\vec{S}(\tau') \rule[-1.6ex]{0ex}{6ex} \right] \nonumber\\
&& +\int_0^{\beta} \! d\tau \, J \vec{S}(\tau) \cdot \vec{s}_{c}(\tau)
+ S_{\text{top}} .
\label{S-imp-kondo-lattice}
\end{eqnarray}
Here $S_{\text{top}}$ describes the Berry phase of the impurity spin.
The Weiss fields $G_0^{-1}$ and $\chi_{0}^{-1}$ are determined
self-consistently from the local lattice Green's function $G_{\text{loc}}$ and
the local lattice spin susceptibility $\chi_{\text{loc}}$, respectively.

The effective action of Eq.~(\ref{S-imp-kondo-lattice}) can be
reformulated in terms of an impurity Hamiltonian,
\begin{eqnarray}
{\cal H}_{\text{imp}}
&=& \sum_{{\bf k},\sigma} E_{\bf k}c_{{\bf k}\sigma}^{\dagger}c_{{\bf k}\sigma}
+ \sum_{\bf q} w_{\bf q}\,\vec{\phi}_{\bf q}^{\;\dagger}\cdot\vec{\phi}_{\bf q}
\nonumber\\ &&
+ \; J \vec{S} \cdot \vec{s}_{c} + g \sum_{\bf q} \vec{S} \cdot
\left( \vec{\phi}_{\bf q} + \vec{\phi}_{-{\bf q}}^{\;\dagger} \right) ,
\label{H-imp-kondo-lattice}
\end{eqnarray}
where the effective dispersion $E_{\bf k}$ is determined by $G_0^{-1}$,
while $w_{\bf q}$ and $g$ are determined by $\chi_{0}^{-1}$.
The conduction band describes at the single-particle level the effect on the
selected site of all electrons at other sites in the lattice.
The vector boson fields $\vec{\phi}_{\bf q}$
extend this description to the particle-hole (spin)
level by keeping track of nonlocal quantum fluctuations.
These fields are the dynamical analogs of the Weiss field
in the mean-field treatment of classical spin systems.

Just as in the standard large-$D$ limit, once the local problem is solved for
the impurity self-energy $\Sigma_{\text{imp}}(\omega)$, one can calculate the
lattice Green's function via the relation
\begin{equation}
G ({\bf k}, \omega) = \frac {1}  {\omega + \mu - \epsilon_{\bf k}
- \Sigma_{\text{imp}}(\omega)} ,
\label{G_k}
\end{equation}
where $\epsilon_{\bf k}$ describes the dispersion on the physical
$D$-dimensional lattice.
The lattice spin susceptibility is more complicated to determine
because it satisfies an integral (Bethe-Salpeter) equation
rather than an algebraic (Dyson) equation. In the extended DMFT,
however, the momentum-dependent spin susceptibility is still
relatively simple:
\begin{equation}
\chi ({\bf q}, \omega) ={1 \over  {M (\omega) - I({\bf q}) }} .
\label{chi_q_cumulant}
\end{equation}
Here $I({\bf q})$ is the Fourier transform of $I_{ij}$,
while $M(\omega)$, reflecting the local dynamics,
is given by the Weiss field $\chi_0^{-1}$ and the
local spin susceptibility $\chi_{\text{loc}}$ as follows:
\begin{equation}
M(\omega) = \chi_{0}^{-1}(\omega) + {1 \over \chi_{\text{loc}}(\omega)} .
\label{eff_cumulant}
\end{equation}
Both $\chi_0^{-1}$ and $\chi_{\text{loc}}$ are entirely determined by
the effective impurity problem. The detailed diagrammatic
derivations of Eqs.~(\ref{chi_q_cumulant}) and~(\ref{eff_cumulant})
can be found in Ref.~\onlinecite{Smith3}.

A zero-temperature phase transition will occur in the Kondo lattice
when $\chi ({\bf Q}, \omega=0)$ becomes divergent for some momentum ${\bf Q}$.
In the extended DMFT formulation of the problem, $M(\omega)$ may be regular
at the transition point [i.e., $\text{Im}\, M(\omega)\sim \omega$], in which
case the quantum phase transition will be of the usual spin-density-wave
(SDW) type.
However, should $M(\omega)$ instead have an anomalous frequency dependence at
the transition, then the local critical behavior will also be
anomalous.

In order to distinguish between the two types of transition, one must
overcome the significant hurdle of solving a self-consistent Kondo problem
with both a fermionic band and a vector bosonic bath.
Progress in the standard large-$D$ DMFT was built on several decades'
accumulation of knowledge about the conventional Kondo and Anderson
impurity models. Similar advances will be made within the extended DMFT
only once the novel quantum impurity problem is understood.
The next section outlines preliminary steps in this direction through
analysis of the impurity problem without the imposition of self-consistency.

\section{Quantum Phase Transitions in Generalized Kondo Problems}
\label{Kondo}

Within the extended DMFT framework described in
Section~{\ref{mean-field-theory}, the nature of a quantum phase transition
arising in the Kondo lattice can be deduced from the frequency dependence of
the quantity $M(\omega)$.
Local Kondo physics will dominate if $M(\omega)$ is anomalous
at the transition.
Since it is entirely determined by the effective impurity problem
[Eq.~(\ref{H-imp-kondo-lattice})], $M(\omega)$ can be anomalous only
if the impurity problem has its own critical point with an anomalous
local susceptibility.
At the same time, self-consistency dictates that the spectral functions
of the fermionic band and the vector bosonic bath must also be anomalous.
Rather than determining the forms of these spectral functions
using the extended DMFT self-consistency conditions, we have instead
{\em assumed} certain unconventional forms from the outset.
We show in this section that such forms can give rise to quantum critical
points with local susceptibility anomalies.

\subsection{\bf Kondo problem with an additional bosonic bath}
\label{Kondo+boson}

We first consider the Kondo impurity problem given in
Eq.~(\ref{H-imp-kondo-lattice}).
Below some high-energy cutoff scale $1/\xi_0$,
we assume that the conduction electron density of states is a
(nonzero) constant,
\begin{equation}
\rho(\epsilon) \equiv \sum_{\bf k} \delta (\epsilon - \epsilon_k)
= \rho_0 ,
\label{dos-fermion}
\end{equation}
and take the spectral function of the bosonic bath to be of the form
\begin{equation}
\sum_{\bf q} \delta (\omega - w_{\bf q}) = (K/\xi_0)
(\omega \xi_0)^ A .
\label{dos-boson}
\end{equation}
In the self-consistent problem, $A$ would be determined from the form of the
local spin susceptibility $\chi_{\text{loc}}$.
For a Fermi liquid, $\text{Im}\,\chi_{\text{loc}}(\omega) \sim \omega$
or equivalently, $\chi_{\text{loc}} (t) \sim 1/ t^2$; as a result,
the corresponding boson spectral function would be Ohmic, i.e., $A=1$.
In a non-Fermi liquid for which $\chi_{\text{loc}}
(t) \sim 1/ t^{\alpha}$ with $\alpha\not= 2$, the boson spectral
function would be non-Ohmic, i.e., $A = \alpha -1 \ne 1$.
Here, we simply assume that $A$ takes some fixed input value.

The Kondo impurity problem described by Eqs.~(\ref{H-imp-kondo-lattice}),
(\ref{dos-fermion}), and~(\ref{dos-boson}) has recently been studied
via a $1-A$ expansion.\cite{Smith2,Sengupta}
Upon progressive reduction of the high-energy cutoff, the couplings
$J$ and $g$ entering Eq.~(\ref{H-imp-kondo-lattice}) satisfy the
renormalization-group (RG) equations
\begin{eqnarray}
{d \ln J} \over {d \ln \xi}
&=& (\rho_0 J) - {1 \over 2} (\rho_0 J) ^2
- K g^2 ,  \nonumber\\
{d \ln g }\over{d ln \xi} &=& 1-A -  {1 \over 2} (\rho_0 J)^2 - K g^2 .
\label{scaling.gJ}
\end{eqnarray}

\begin{figure}
\centering
\vbox{\epsfysize=55mm\epsfbox{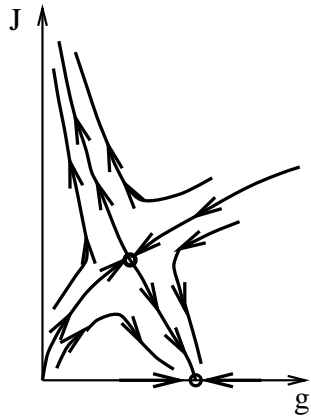}}
\vspace{1ex}
\caption{Schematic renormalization-group flows for the Kondo problem with a
additional bosonic bath described by an exponent $A<1$.}
\label{rg-flow-kondo-boson}
\end{figure}

The RG flows for $A\lesssim 1$ are shown in
Fig.~\ref{rg-flow-kondo-boson}. For small $J$ and small $g$,
there exists a separatrix specified by $ \sqrt{K} g_{c} = (1-A)
\exp[-(1-A)/(\rho_0J)]$.
The Kondo coupling $J$ flows towards strong coupling
for $g < g_{c} $ and towards zero for $g > g_{c}$.
More generally, the flows are controlled by an unstable fixed
point located at $(\sqrt{K}g^*, \rho_0J^{*})
= (\sqrt{1-A}, 1-A)$.

Within the $1-A$ expansion, it is found the local spin correlation function
at the critical point has an anomalous exponent\cite{Sengupta}:
$\chi_{\text{loc}}( t ) \sim 1/t^{\beta}$, where $\beta=1-A$.

\subsection{\bf Kondo problem with a pseudo-gap}
\label{Kondo+pseudo-gap}

In the previous subsection, the fermionic spectral function
was assumed to be regular.
It is possible, however, that critical fluctuations may produce
a pseudogap in the effective single-particle density of states
of the lattice model and, hence,
in the density of states of the fermionic band of
the impurity model.
In order to isolate the effects of such a pseudogap, we neglect the
bosonic bath in this subsection.
Specifically, we study the SU($N$)-symmetric Kondo impurity
Hamiltonian
\begin{equation}
{\cal H}_{\text{imp}} = \sum_{{\bf k},\sigma}
E_{\bf k}
    c_{{\bf k}\sigma}^{\dagger} c_{{\bf k}\sigma}
  + {J \over N} \sum_{\sigma,\sigma'} \vec{S} \cdot c_{0\sigma}^{\dagger}
    \vec{\tau}_{\sigma \sigma'} c_{0\sigma'}.
\label{Kondo.hamiltonian}
\end{equation}
Here $\vec{S}$ is an impurity moment of spin degeneracy $N$,
$c_{0\sigma}$ ($\sigma = 1, 2, \ldots, N$)
annihilates an electron at the impurity site,
and $\tau^i_{\sigma \sigma'}$ ($i=1, 2, \ldots, N^2\!-\!1$) is
a generator of SU($N$).
The conduction band density of states is
assumed to take the power-law form
\begin{equation}
\rho(\epsilon) = \left\{
   \begin{array}{ll}
       \rho_0 |\epsilon|^r \quad & \text{for } |\epsilon| \le 1, \\[0.5ex]
       0 & \text{for } |\epsilon| > 1.
   \end{array} \right.
\label{dos}
\end{equation}

The problem described by Eqs.~(\ref{Kondo.hamiltonian}) and~(\ref{dos}) was
first studied by Withoff and Fradkin.\cite{Withoff}
The local moment is quenched at low temperatures only if the Kondo
coupling exceeds a threshold value $J_c$, where $\rho_0 J_c \approx r$.
This makes the critical point at $J=J_c$ a genuinely interacting problem, in
sharp contrast to the conventional ($r=0$) Kondo model, for which $J_c=0$.
The strong-coupling (i.e., $J>J_c$) and weak-coupling ($J<J_c$)
properties of the power-law problem have
been studied extensively.\cite{Cassanello,Chen,Gonzalez-Buxton}

In a recent work,\cite{Ingersent} we studied the quantum critical behavior
of the power-law Kondo model. In this context, it turns out to be crucial
to distinguish between two different spin susceptibilities:
the thermodynamic quantity $\chi_{\text{imp}}$, which measures the
response to a magnetic field that couples both to the impurity spin and to
the conduction electron spins, minus the equivalent response in the absence of
the impurity;
and the local susceptibility $\chi_{\text{loc}}$, the response to a magnetic
field which couples only to the impurity spin.
In the conventional Kondo problem, $\chi_{\text{loc}}$ closely tracks
$\chi_{\text{imp}}$ as a function of temperature.\cite{Chen:92}
For a power law density of states, by contrast, $\chi_{\text{loc}}$ and
$\chi_{\text{imp}}$ turn out to behave very
differently\cite{Chen,Gonzalez-Buxton} as the Kondo coupling
approaches $J_c$. Only $\chi_{\text{loc}}$
exhibits the scaling behavior characteristic of a continuous phase transition.

\begin{figure}
\centering
\vbox{\epsfxsize=65mm\epsfbox{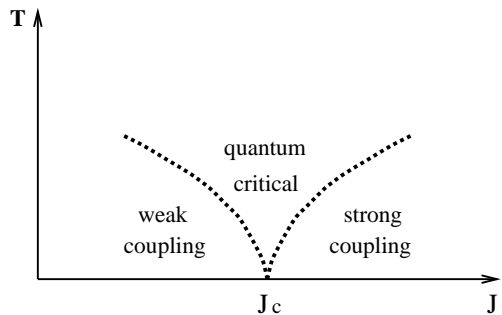}}
\vspace{1ex}
\caption{Schematic phase diagram of the Kondo problem with
a power-law pseudogap.}
\label{pseudogap}
\end{figure}

This insight allows us to identify the form of the singular part
of the free energy, from which we derive scaling relations between the
critical exponents in the quantum critical, weak-coupling, and
strong-coupling regimes (see Fig.~\ref{pseudogap}).
Of particular interest is the exponent $x$ which governs the
temperature-dependence of the local susceptibility at the critical point via
the relation $\chi_{\text{loc}}(T, J = J_c) \sim 1/T^x$.
As outlined below, $x$ can be calculated in three separate limits.

(1) For $N\rightarrow\infty$, the exponents on the strong-coupling side
of the transition ($J>J_c$) can be calculated using slave-boson mean-field
theory.
Substituting these exponents into the scaling relations leads to the
prediction that $x(N=\infty)=1$, independent of $r$.

(2) When $r$ is small, $\rho_0 J_c$ is also small, and $x$ can be calculated
directly via a perturbative expansion in $\rho_0 J_c$.
The result is
\begin{equation}
x(\rho_0 J_c \ll 1) = 1 - \frac{2}{N}(\rho_0 J_c)^2 .
\label{x:small_r}
\end{equation}

(3) For $N=2$, we have calculated the critical exponents
using the numerical RG method. It is found that
$x$ varies continuously with $r$, and obeys
Eq.~(\ref{x:small_r}) closely even for $r$ as large as $0.45$.

The three independent methods for calculating $x$ produce results which are
consistent with one another and point to the conclusion that for any finite
$N$, $x$ varies continuously with $r$.

In summary, we have studied two kinds of generalized Kondo impurity problems
featuring unconventional host media. In both cases, there arises a quantum
critical point with anomalous dynamics, such that the
$T=0$ dynamical susceptibility varies as
\begin{equation}
\text{Im}\,{\chi}_{\text{loc}}(\omega) \sim \omega^B ,
\label{chi-imp}
\end{equation}
where $B$ is less than $1$ and depends on
the form of either the boson spectral function or
the fermion pseudogap.

\section{Quantum Phase Transitions in Heavy Fermions}
\label{qpt}

The results for generalized Kondo impurity problems reviewed above,
when combined with the extended DMFT formalism summarized
in Section \ref{mean-field-theory}, suggest that it is indeed
possible for a Kondo lattice system to display novel quantum
critical behavior with anomalous local dynamics. In this
section we expand on this point. The arguments
of this section, unlike those of the previous two sections,
are necessarily suggestive in nature.

We will find it useful to define an energy scale $E_{\text{loc}}^*$
below which $M(\omega)$ defined in Eq.~(\ref{eff_cumulant}) is
regular. Specifically,
\begin{equation}
\text{Im}\,M (\omega) = \text{const.} \, {\omega \over [E_{\text{loc}}^*]^2}
\qquad \text{for } \omega \ll E_{\text{loc}}^*,
\label{E-loc-*}
\end{equation}
where the constant is of order unity.

For small values of the tuning parameter $\delta$, the system lies well inside
the paramagnetic region of Fig.~\ref{qpt-hf}. In such cases, the RKKY
coupling $I_{RKKY}$ is negligible, and the mean-field equations of the
extended DMFT reduce to those of the standard DMFT.\cite{Georges}
The effective impurity problem is a Kondo model with a regular
conduction electron band, for which both the local susceptibility
$\chi_{\text{loc}}$ and the Weiss field $\chi_0^{-1}$ have
the usual Fermi liquid form.
Equation~(\ref{eff_cumulant}) then implies that
$\text{Im}\,M(\omega)\sim\omega$, i.e., that $E_{\text{loc}}^*$ is finite.

Now let us consider larger values of $\delta$, for which the RKKY
interaction becomes more significant. Let us suppose that the extended
DMFT for the Kondo lattice self-consistently generates an effective
impurity problem with an unconventional fermionic or bosonic
bath of the type discussed in the previous section.
Varying the value of $\delta$ will change the couplings in the
impurity problem.
The results of Section~III suggest that for some
choice $\delta = \delta_{\text{loc}}^c$, the impurity model will be
tuned to a critical point characterized by a local susceptibility of
the form given in Eq.~(\ref{chi-imp}).
This in turn implies that at the critical point
$\text{Im}\,M(\omega) \sim \omega ^B$ with $B<1$, and hence
that $E_{\text{loc}}^* = 0$.
We infer that $E_{\text{loc}}^*$ must vary with $\delta$ roughly as
shown in Fig.~\ref{e-loc}.

We can determine whether anomalous local physics plays an important role in
the lattice quantum critical behavior by comparing $\delta_{\text{loc}}^c$
defined above with $\delta_c$, the smallest value of $\delta$ for which the
momentum-dependent spin susceptibility $\chi({\bf Q},\omega=0)$ diverges at
any wavevector ${\bf Q}$.

If $\delta_{c}$ is significantly smaller than $\delta_{\text{loc}}^c$,
then $E_{\text{loc}}^*$ remains finite (and acts as the effective Fermi
energy) throughout the paramagnetic region right up to the critical point.
In this case, the magnetic phase transition represents
a Fermi-surface instability induced by the RKKY
interaction.\cite{two-stage} This is similar to the transition
in more weakly interacting electron systems.
The Hertz-Millis approach\cite{Hertz,Millis} should provide an appropriate
basis for understanding the quantum critical behavior.

Alternatively, $\delta_{c}$ may be equal (or very close) to
$\delta_{\text{loc}}^c$, in which case $E_{\text{loc}}^*$ vanishes
(or nearly vanishes) at the critical point.
The quantum critical behavior in this case is dictated by the anomalous
local dynamics.
Equations~(\ref{chi_q_cumulant}) and~(\ref{eff_cumulant}) imply that
$\chi({\bf q},\omega)$ will be anomalous over the entire Brillouin zone.

\begin{figure}
\centering
\vbox{\epsfxsize=65mm\epsfbox{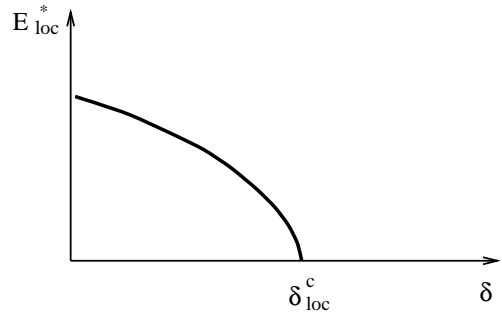}}
\vspace{1ex}
\caption{The local energy scale defined in Eq.~(\protect\ref{E-loc-*})
as a function of the tuning parameter.}
\label{e-loc}
\end{figure}

The second type of critical point represents a continuous
transition from a paramagnetic metal, having a Fermi surface
of volume $N_f+N_c$, to a magnetic metal, with a Fermi surface
of volume $N_c$. Here $N_f$ and $N_c$ denote the number of
$f$-electron spins and conduction electrons, respectively.
One can think of the anomalous local dynamics as capturing
the fluctuations between the two Fermi surfaces.
(This last picture has been noted previously.\cite{Varma})

The discussion to this point has focused on the intermediate quantity,
$M (\omega)$.
However, the measurable quantity in a Kondo lattice system
is $\chi({\bf q},\omega )$.
From Eq.~(\ref{chi_q_cumulant}) it is apparent that
the low-frequency behavior of $\text{Im}\,M (\omega)$ is the
same as that of $\text{Im}\,\tilde{\chi}_{\text{loc}}$, where
\begin{equation}
\tilde{\chi}_{\text{loc}} = \sum_{\text{generic}~{\bf q}}
\chi ({\bf q}, \omega) .
\label{chi-generic}
\end{equation}
Here ``generic ${\bf q}$'' restricts the summation to wavevectors that
are not too close\cite{generic} to the ordering wavevector ${\bf Q}$.

The preceding discussion suggests an operational criterion
for distinguishing the two types of critical behavior.
Let us redefine the energy scale $E_{\text{loc}}^*$ so that
\begin{equation}
\text{Im}\,\tilde{\chi}_{\text{loc}} \sim \omega \qquad
\text{for } \omega \ll E_{\text{loc}}^* .
\label{E-loc-*-2}
\end{equation}
Then, if $E_{\text{loc}}^* =0$ at the quantum critical point, the
phase transition is driven by local physics.
Otherwise, the transition is of the usual SDW type.

Neutron scattering experiments have recently been
carried out\cite{Schroder,Stockert}
on CeCu$_{6-x}$Au$_x$ at the critical
concentration $x_c=0.1$.
At wavevectors close to the ordering wavevectors,
the dynamical spin susceptibility appears to have
an anomalous frequency dependence,
with an exponent $B \approx 0.8$.
Furthermore, the same exponent also describes the
${\bf q}=0$ susceptibility, suggesting\cite{Schroder}
that the anomalous energy exponent occurs over a large
region of the Brillouin zone. The latter observation,
according to Eqs.~(\ref{chi-generic}) and~(\ref{E-loc-*-2}),
implies that $E_{\text{loc}}^* =0$ at the critical point.
Thus, the quantum critical behavior of CeCu$_{6-x}$Au$_x$
appears to fall in the category of quantum phase transitions
driven by local physics.

\section{Summary and Outlook}
\label{sec:conclu}

To summarize, we have reviewed an extended DMFT which maps the Kondo
lattice problem onto a generalized Kondo impurity problem, supplemented
by self-consistency conditions. The Kondo and RKKY physics
of the lattice model are both manifested in the effective
impurity problem: The former is described by the coupling
of the local moment to a self-consistent conduction electron
band and the latter by the coupling of the local moment to
a self-consistent bosonic bath.

Two kinds of quantum phase transitions are argued to occur
in Kondo lattice systems. One is the usual SDW type,
whereas the other is governed by anomalous local Kondo dynamics.
For quantum critical behavior of the second type to occur,
it is necessary that the generalized Kondo impurity arising in
the extended DMFT displays a critical point at which
the local susceptibility has an anomalous exponent.
Such critical points occur in two related
Kondo impurity models, one describing a local moment
coupled to an additional bosonic bath, the other
featuring a pseudogap in the conduction band.

Finally, we have introduced an operational criterion that
can be used to identify quantum critical behavior
driven by local physics. It is based on the vanishing
of a local energy scale [$E_{\text{loc}}^*$ defined in
Eqs.~(\ref{chi-generic}) and~(\ref{E-loc-*-2})] at the critical point.
The results of neutron scattering experiments\cite{Schroder}
suggest that this criterion is satisfied in CeCu$_{6-x}$Au$_x$.

A number of significant issues remain to be addressed.
Most importantly, the self-consistent mean-field equations
need to be analyzed more completely. Once that has been
achieved, it should be possible to study the effect of disorder
on this novel type of quantum phase transition.\cite{disorder}
Finally, the effect of valence fluctuations can also be
addressed within this framework.

We would like to thank P.\ Coleman, G.\ Kotliar,
A.\ Rosch, S.\ Sachdev, A.\ Sengupta,
and C.\ M.\ Varma for useful discussions.
This work has been supported in part by
NSF Grant No. DMR--9712626, Research Corporation,
and the A.\ P.\ Sloan Foundation.


\begin{references}
\bibitem{two-stage}
A second transition (or crossover) can in principle
take place on the ordered side.
\end{references}
\end{document}